# Per-Span Microwave Frequency Fiber Interferometry in Subsea Cables for Scalable Deep-Ocean Geophysical Monitoring

Georgios Aias Karydis[1], Nicolas L. Celli[2], David Craig[2], Örn Jónsson[3], Andrés Arnar Hlynsson[3], Eoin Kenny[4], Charis Mesaritakis[5], Christopher J. Bean[2] and Adonis Bogris[1]

(1)Dept. of Informatics and Computer Engineering, University of West Attica, Egaleo, Greece gakarydis@uniwa.gr, abogris@uniwa.gr
(2)Dublin Institute for Advanced Studies, School of Cosmic Physics, Geophysics Section, Ireland
(3)Farice, Gudridarstigur 2-4, 113, Reykjavik, Iceland
(4)Asiera CLG, Dublin
(5)Dept. of Biomedical Engineering, University of West Attica, Egaleo, Greece.

**Abstract** *We demonstrate low-cost microwave frequency fibre interferometry for per-span monitoring of a 1,770 km operational subsea cable connecting Ireland and Iceland. Over four months, this approach successfully resolved tidal variations, storms, and teleseismic earthquakes, highlighting its potential for large-scale, cost-effective ocean monitoring.* 

## Introduction

Geophysical monitoring via fibre optic sensing offers a powerful new opportunity, enabled by the global network of submarine cables that provide a sensing platform in remote areas where traditional instruments are difficult to deploy. Distributed Acoustic Sensing (DAS) is currently the most mature and high-resolution fibre interrogation technique. However, its application in long-haul, traffic-carrying submarine cables remains challenging, even with recent pioneering experiments demonstrating its potential on operational subsea networks [1, 2]. G. Marra at al. [3] presented a practical pioneering route towards the interrogation of long-haul submarine links by leveraging the existing high-loss loop backs (HLLBs). HLLBs transform the single long link into an array of sensors at a resolution dictated by the periodic placement of optical amplifiers across the link. The link provides localization capabilities at a resolution of 50-100 km which is adequate for the detection and localization of geophysical effects such as tele-seismic earthquakes, ocean swells and tsunamis [3]. Since then, there have been a few more demonstrations relying on per span optical frequency domain reflectometry (OFDR) [4] polarization measurements [5]. Polarization based sensing presented in [5] is very practical but has limited sensitivity compared to phase interrogation techniques. OFDR based per-span interferometry is really efficient and has been shown to be capable of global seismic monitoring [4], however it requires a high-quality fibre laser, high performance and high-speed digital electronics (~ Gsps) and GPU processing which increase the cost of the interrogator —limiting the application of the technology on multiple cables globally— and make real-time signal processing more challenging.

A technique compatible with long-haul cable interrogation is the so-called microwave frequency fiber interferometry (MFFI) [6]. The theoretical framework and the effectiveness of MFFI for earthquake detection have been demonstrated in both terrestrial and submarine installations [6–7]. This technique offers distinct advantages in terms of simplicity and cost-efficiency, as it replaces high-quality, low-linewidth lasers with off-the-shelf 10 GHz oscillators exhibiting superior absolute spectral purity (sub-Hz linewidth). This substitution enables a low-noise, sub-mHz response that is essential for geophysical monitoring. More recently, a digital variant of MFFI [8] enabled distributed measurements of tides and very-long-period temperature variations over a transatlantic cable connecting Portugal and Brazil, combining phase measurements in the digital domain using a 90 MHz microwave modulation with digital downconversion of phase variations. This approach limits the sensitivity to sub-millihertz measurements—below the frequency range of most oceanographic and seismic signals—and is very demanding in terms of hardware and data processing at frequencies offering higher sensitivity in the GHz range. Here, we report the first implementation of a 10 GHz MFFI with analogue downconversion in a per-span interferometric configuration. We demonstrate its ability to sample a broad frequency range (hours to 25 Hz), and detect not only ocean tides, but also ocean storms and teleseismic earthquakes with high fidelity at low sampling rates (<Msps range) simplifying data acquisition processes. These results establish a foundation for a low-cost fibre sensing approach that supports sensitive, real-time data processing for early warning and live monitoring applications.

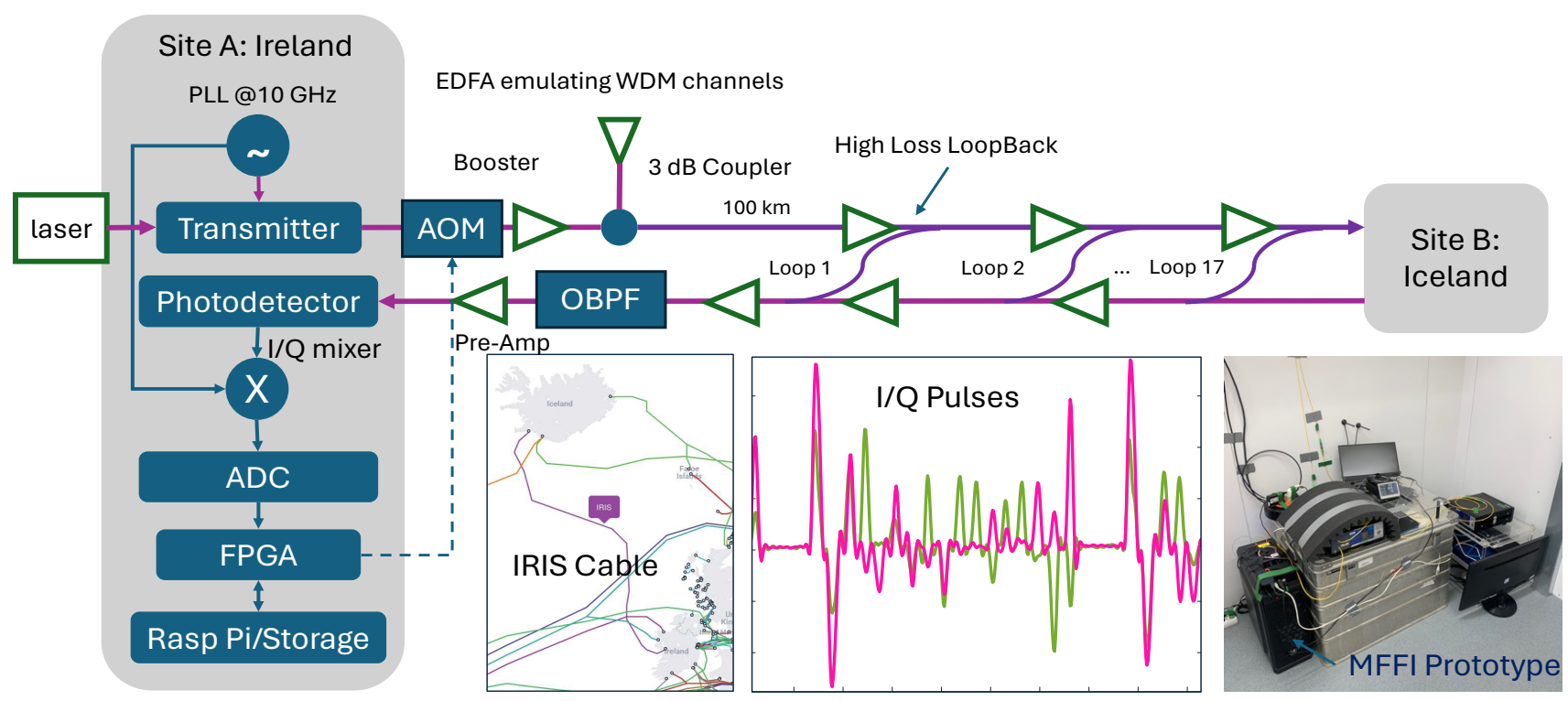

**Fig. 1:** Experimental testbed and prototype description.

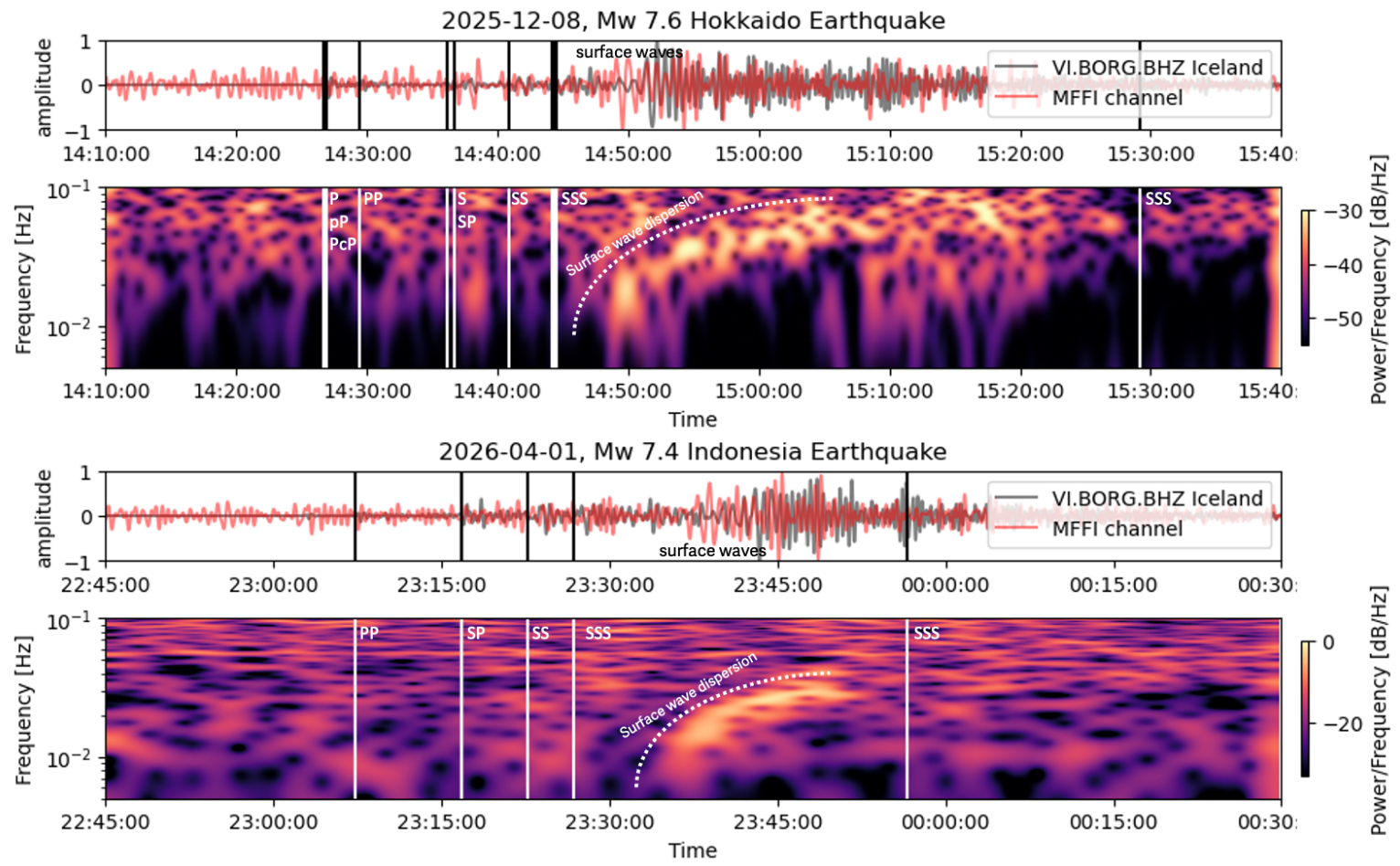

**Fig. 2**: Teleseismic earthquakes from Japan (top) and Indonesia (bottom) recorded on the MFFI channel 17, shown as time traces and spectrograms.

## Prototype and experimental testbed

The experimental testbed is depicted in fig. 1. The MFFI uses a tunable laser adjusted to the HLLB wavelength (~ 1565 nm in our case). The laser enters the prototype which comprises a Mach-Zehnder modulator (MZM) directly driven by a phase-locked loop (PLL) to modulate the optical intensity at 10 GHz. The IRIS cable is about 1770 km long (~ 17.5 ms round trip) connecting Iceland to Ireland (see fig.1). The cable is operated by Farice and the MFFI unit is placed at the cable landing station in Galway, Ireland. The output of the MZM is driven to an acousto-optic modulator (AOM) in order to apply 0.5 ms pulses (~ 1 ms inter-span delay – 100 km amplifier distance) for gating purposes at a repetition period of 20 ms (~ 17.5 ms round-trip delay). The MFFI output is first amplified by a booster and combined with amplified spontaneous emission noise to emulate multi-wavelength transmission and stabilize the in-line amplifiers. After propagation, the returned signal, degraded by HLLB losses and accumulated noise, is bandpass filtered (OBPF in Fig. 1) and pre-amplified. It is then detected by a 10 GHz photodetector and down-converted to baseband using an analogue I/Q RF mixer. The I/Q signals are electrically amplified, low-pass filtered, and sampled at 1 Msps by a commercial ADC. Data acquisition and decimation (×60, to ~16 ksps) are performed on an FPGA. The FPGA also applies the pulses to the AOM, while a Raspberry Pi handles system control and data storage. The PLL was tuned to 9.8 GHz to mitigate chromatic-dispersion-induced fading across the 17 loops. The recovered I/Q pulses (20 ms period) shown in Fig. 1 demonstrate adequate Signal to Noise Ratio (SNR) for all loops without any compensation. The prototype unit is shown in Fig. 1 (black box).

## Results and Discussion

The MFFI unit has been interrogating the cable since November 2025. In this period, it successfully demonstrated its ability to detect a large variety of geophysical processes in the ocean. The first example is the detection of tele-seismic waves generated by the 08/12/2025 Mw 7.6 Hokkaido earthquake and 01/04/2026 Mw 7.4 Indonesia earthquake. Fig. 2 shows that, for both earthquakes, surface waves are clearly detected and there is an indication of S and SSS waves as well. These traces were captured in the 17th span of the link (close to Iceland), but similar traces were detected in many other channels/spans. For both teleseisms, spans closest to Iceland have the best SNR, showing partial agreement with seismic stations in Iceland especially for surface waves (differences are expected and due to a number of factors, e.g.: different sensed physical quantities; spatially integrated- vs. point measurements; location; different geometrical sensitivity). The time trace of the MFFI is displayed in red and is compared to the velocity recordings from the BORG seismic station in Iceland. Theoretical arrival times for seismic waves are superimposed on both panels and labelled. Surface wave dispersion is highlighted in the spectrogram.

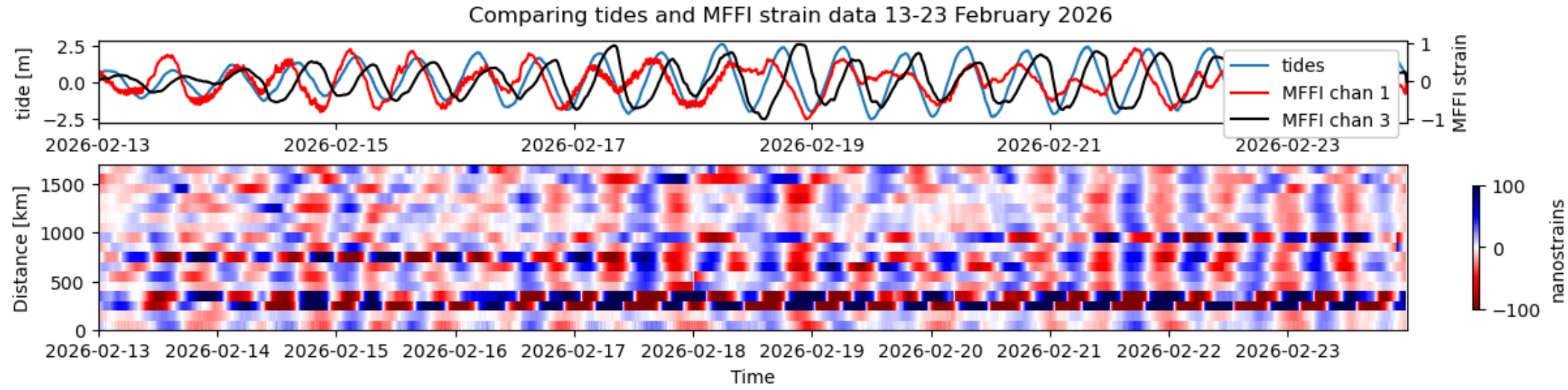

**Fig. 3**: Ocean tidal effects in the North East Atlantic recorded by the MFFI.

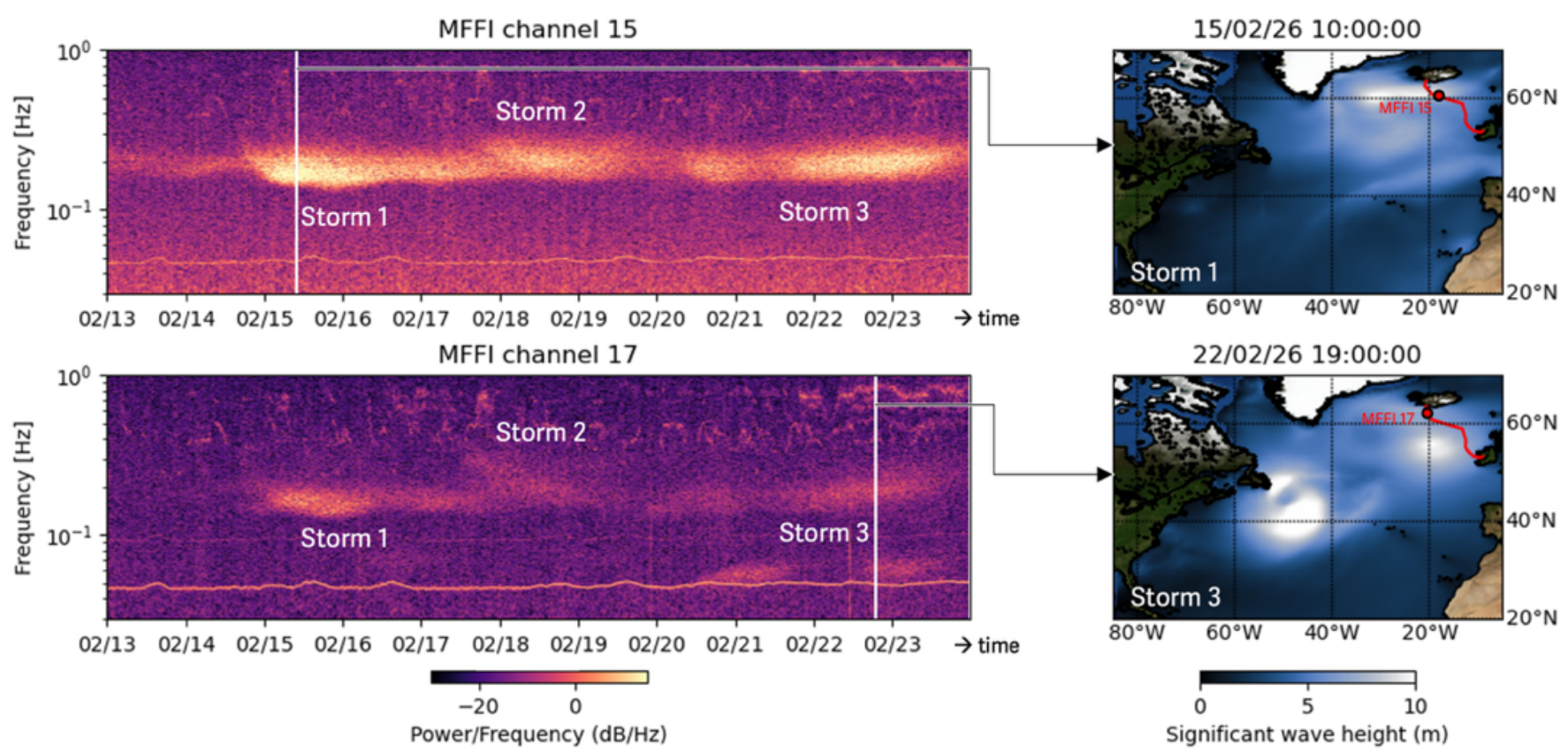

**Fig. 4**: February 2026 ocean storms sensed by the MFFI.

We also measured, for the first time, tidal effects across the whole North East Atlantic as depicted in fig. 3. All spans (filtered 0.00001-1 Hz) record the tidal signal clearly (bottom panel, with channels sorted by their distance from Galway), and a comparison with real ocean tide height measured at Galway Port (blue) to channels 1, 3 of the MFFI (red and black, respectively) with patterns showing excellent agreement. The tidal signal is also strongest on spans 3, 4 and 8, which are closest to steep bathymetric gradients in the Rockall basin, which can suggest stronger sea-bottom currents.

Finally, we successfully recorded ocean storms across the whole length of the cable, with different spans showing local variations in the recorded signals (fig. 4). The spectrograms show examples of storm-induced microseisms (0.1-0.3 Hz) recorded by MFFI channels 15, 17 between 13-23/02/2026. Panels to the right show the significant ocean wave height from the WAVEWATCH III hindcast data for the storms on the 15th and 22nd of Feb. 2026, indicating a clear correlation between the ocean activity and recorded microseisms.

So far, we have observed a broad spectral response across several geophysical processes, including ocean tides (~12-hour period, figure 3), two teleseismic earthquakes (~15–50 s period, figure 2), and ocean-generated secondary microseisms (~5 s period, figure 4). Teleseismic earthquake observations help fill gaps in global seismic data coverage, potentially enabling improved imaging of the Earth's deep interior [9]. Observations of ocean microseisms and tides are arguably even more compelling, as they open opportunities for direct, near-real time, deep-ocean monitoring, including climate-related changes in the water column [10] and the stability of offshore deep-water slopes. Our demonstrated ability to measure offshore (>100 km from coastlines) long-period ocean-floor pressure fluctuations may also offer significant potential for global tsunami early-warning systems.

## Conclusions

We present per-span MFFI results from a subsea cable linking Iceland and Ireland, demonstrating sensitivity to key oceanic and geological processes across a broad spectral range. Affordable components such as low-cost PLLs at 10 GHz and minimal signal processing enable robust real-time detection and classification revealing that per-span MFFI can be instrumental for the large-scale sensing of multiple cables.

## Acknowledgements

We would like to thank Mikael Mazur and Thomas Nikas for fruitful discussions. This work has been partially funded by Horizon Europe ECSTATIC under Grant Agreement 101189595 and by Research Ireland Infrastructure and Pathway awards.